\documentclass[preprint,12pt]{elsarticle}
\usepackage{amsmath,amssymb,amsfonts,amsthm,mathtools}
\usepackage{bm}
\usepackage{physics}
\usepackage{placeins}
\usepackage{booktabs}
\usepackage{multirow}
\usepackage{enumitem}
\usepackage{xcolor}
\usepackage{subcaption}
\usepackage{hyperref}
\usepackage{cleveref}
\hypersetup{colorlinks=true, linkcolor=blue, citecolor=blue, urlcolor=blue}
\newcommand{\R}{\mathbb{R}}

\newcommand{\cG}{\mathcal{G}}

\begin{document}
\biboptions{numbers,sort&compress}

\begin{frontmatter}

\title{Neural-Operator-Predicted Time-Dependent Reduced Subspaces for Projection-Based Simulation of Nonlinear PDEs}

\author[inst1]{Boxi Song\corref{cor1}}
\ead{songor@bc.edu}

\author[inst2]{Rennie Mirollo}

\author[inst1]{Jan R.~Engelbrecht}

\cortext[cor1]{Corresponding author.}
\affiliation[inst1]{organization={Department of Physics, Boston College}, 
city={Chestnut Hill}, postcode={02467}, state={MA}, country={USA}
}
\affiliation[inst2]{organization={Department of Mathematics, Boston College}, 
city={Chestnut Hill}, postcode={02467}, state={MA}, country={USA}
}

\begin{abstract}
Accurate numerical simulation of nonlinear partial differential equation initial value problems typically requires high-dimensional full-order discretizations at substantial computational cost. While purely data-driven surrogates can speed up prediction, these "black box" approximations do not necessarily satisfy the governing equations during inference. In this paper we present a hybrid framework, which combines neural operators with projection-based reduced-order modeling for nonlinear PDE simulation. The key idea is to train a neural operator to map the initial condition to a time-dependent reduced subspace, and then evolve low-dimensional reduced coordinates by solving the governing dynamical equations projected on this learned moving trial subspace. In this way, the network predicts an adaptive reduced representation rather than the full trajectory directly, while the online solver preserves a physics-based reduced evolution.

We test this method on two representative nonlinear PDEs: the viscous Burgers' equation and the Fisher--KPP reaction--diffusion equation. In both cases, the proposed approach achieves successful reduced-order simulation with substantial wall-clock speedup relative to the corresponding full-order solver while maintaining relative errors on the order of \(10^{-2}\) to \(10^{-1}\) across the tested resolutions. These results demonstrate that a low-dimensional learned time-dependent reduced subspace can be used to predict solutions to nonlinear PDEs while retaining the advantages of projection-based reduced dynamics. Overall, this method provides a promising framework for fast and physically grounded data-assisted simulation of nonlinear PDEs.
\end{abstract}

\end{frontmatter}

\section{Introduction}

Direct numerical simulation of nonlinear time-dependent partial differential equations (PDEs) is a central computational challenge in fluid dynamics, reaction-diffusion systems, transport phenomena, and scientific computing more broadly. When a discretized PDE state has dimension \(N \gg 1\), repeated simulations for many initial conditions or parameters can be prohibitively expensive. This challenge is especially acute in settings such as design loops, inverse problems, uncertainty quantification, optimal control, and many-query forecasting, where one must repeatedly integrate the same governing equations over a family of inputs.

Projection-based reduced-order modeling (ROM) seeks to reduce this cost by approximating the full-order solution in a low-dimensional subspace, typically constructed offline from solution snapshots. Classical reduced order methods based on proper orthogonal decompositions (POD) and Galerkin or Petrov--Galerkin projections have demonstrated strong effectiveness in many settings and remain a foundational paradigm in model reduction \cite{Sirovich1987,BennerGugercinWillcox2015,HaasdonkOhlberger2008,HesthavenPagliantiniRozza2022}. However, nonlinear transient dynamics often evolve on solution manifolds that are poorly approximated by a single fixed linear subspace. As a result, a static linear subspace may require many modes to accurately resolve evolving coherent structures or nonstationary spatial features, thereby weakening the compression benefit. This limitation is especially prominent for transport-dominated or strongly nonstationary dynamics, where linear reduced subspaces can exhibit slow approximation-error decay \cite{ReissSchulzeSesterhennMehrmann2018,BarnettFarhatMaday2023}.

At the same time, machine learning methods provide powerful alternatives for learning solution operators directly from data. In particular, neural operators such as DeepONet and the Fourier Neural Operator (FNO) have shown that maps between function spaces can be approximated efficiently and with strong empirical performance on PDE problems \cite{LuJinPangKarniadakis2021,LiKovachkiAzizzadenesheli2021,KovachkiLiLiuAzizzadenesheliBhattacharyaStuartAnandkumar2023}. Yet direct surrogate prediction also has limitations: the resulting model may not explicitly preserve the structure of the governing equations, and long-time or out-of-distribution reliability may be difficult to guarantee. This limitation motivates hybrid approaches that combine learned representations with physics-based reduced simulation.

In this work, we develop such a hybrid strategy for nonlinear PDE initial value problems. Rather than directly learning the full solution trajectory, we train a neural operator to predict a time-dependent reduced subspace from the initial condition on which the reduced governing dynamics are solved. Thus, the machine learning component predicts \emph{where} the solution should live, while the reduced solver determines \emph{how} it evolves according to the PDE.

The use of time-dependent subspaces is motivated by the observation that many nonlinear trajectories are more accurately approximated by a moving low-dimensional subspace than by a fixed linear subspace. Classical adaptive and time-dependent model reduction ideas have long recognized this point \cite{PeherstorferWillcox2015,HesthavenPagliantiniRozza2022,SapsisLermusiaux2009}, and recent work has continued to explore time-dependent subspaces in specialized contexts \cite{PatilBabaee2023,DonelloPalkarNaderiDelReyFernandezBabaee2023}. Our contribution is to place this idea in a neural-operator-guided setting for nonlinear PDE initial value problems: the time-varying reduced trial subspace is inferred from the initial condition before the online solve begins.

We demonstrate the method on two canonical nonlinear PDEs: the viscous Burgers' equation and the Fisher--KPP reaction--diffusion equation. In both cases, the proposed framework achieves substantial wall-clock speedup while maintaining final-time mean absolute relative errors on the order of \(10^{-2}\) to \(10^{-1}\) across the tested resolutions. These experiments suggest that learning a time-dependent reduced subspace can provide a useful middle ground between reduced-order simulation and direct black-box trajectory regression.

\smallskip
The main contributions of this paper are:
\vspace{-4pt}
\begin{itemize}[leftmargin=1.5em]
    \item We propose a neural-operator-guided PDE reduced-order modeling framework that can be generalized across different initial condition fields and different spatial resolutions.
    \item We introduce a formulation in which a neural operator predicts a low-dimensional \emph{time-dependent} reduced subspace conditioned on the initial condition.
    
    \item We demonstrate the method on two nonlinear benchmark problems, viscous Burgers' equation and Fisher--KPP equation, and report substantial wall-clock speedup and final-time relative errors on the order of \(10^{-2}\) to \(10^{-1}\).
    \item We position the method relative to classical ROM, direct neural operator surrogates, and recent hybrid ML--ROM approaches, and we identify practical directions for improving robustness and efficiency.
\end{itemize}

\section{Related Work}

\subsection{Projection-based reduced-order modeling}

Projection-based reduced-order modeling approximates a high-dimensional state \(\bm{u}(t) \in \R^N\) by a lower-dimensional representation in a reduced trial subspace. A standard starting point is POD, often implemented through the method of snapshots, which constructs an optimal linear subspace in the \(L^2\) sense from training data \cite{Sirovich1987}. Reduced subspace methods extend this approach to parametrized PDEs with carefully designed offline/online decompositions and have been developed into a mature framework for both stationary and time-dependent problems \cite{BennerGugercinWillcox2015,HaasdonkOhlberger2008,HesthavenPagliantiniRozza2022}.

Our work builds on this projection-based ROM method, but differs in one central respect: the reduced trial subspace is not fixed in advance or adapted online by intrusive updates. Instead, it is predicted from the initial condition by a trained neural operator and then used in a projected reduced simulation.

\subsection{Time-dependent and adaptive reduced spaces}

A static linear reduced subspace is often insufficient for transport-dominated or nonstationary dynamics. This issue has motivated a variety of localized adaptive and time-dependent subspace strategies. Online adaptive subspace updates for nonlinear systems have also been proposed, for example by coupling low-rank updates with hyper-reduction ideas \cite{PeherstorferWillcox2015}. Time-dependent subspace basis methods also appear in dynamically orthogonal and dynamically bi-orthogonal formulations for stochastic PDEs \cite{SapsisLermusiaux2009,PatilBabaee2023,DonelloPalkarNaderiDelReyFernandezBabaee2023}.

Recent work has further explored ROM with explicitly time-dependent subspaces in more specialized contexts, including stochastic PDEs and evolving boundary-condition settings \cite{PatilBabaee2023,DonelloPalkarNaderiDelReyFernandezBabaee2023}. These works reinforce the core premise that moving low-dimensional spaces can significantly improve approximation efficiency when the relevant solution structures translate, deform, or otherwise evolve in ways that are not well-aligned with a fixed subspace. Our method shares this perspective, but differs in that the time-dependent subspace is predicted from the initial condition by a neural operator rather than being evolved by a separate intrusive subspace-dynamics system.

\subsection{Neural operators and operator learning}

Neural operators can be used to learn maps between function spaces rather than maps between vector spaces with fixed finite dimensions. In particular, Fourier Neural Operators (FNOs) provide a highly effective spectral-kernel parameterization that has become a widely used architecture for PDE learning \cite{LiKovachkiAzizzadenesheli2021,KovachkiLiLiuAzizzadenesheliBhattacharyaStuartAnandkumar2023}. These models can approximate families of PDE solution operators and, in many cases, offer a significant improvement in speed over conventional solvers.

Most neural operator work, however, focuses on the direct prediction of solution fields or the one-step / multi-step evolution operators. In that setting, the learned models replace the numerical solvers. Instead, our work uses a neural operator to predict a time-dependent set of vectors,
$ \left\{ \phi_1(t),\phi_2(t),\ldots,\phi_r(t) \right\}$ which generically span an $r$-dimensional trial subspace $\Lambda(t)$. Thus, the network approximates a map from an initial condition to $\Lambda(t)$
\[
u_0(\cdot) \mapsto \Lambda(t)={\rm span} \, \left\{ \phi_1(t),\phi_2(t),\ldots,\phi_r(t) \right\}
\]
rather than directly approximating
\[
u_0(\cdot) \mapsto u(t,\cdot).
\]
This distinction is important: the online dynamics in our method remain governed by a projected version of the PDE.

\subsection{Hybrid machine learning and ROM approaches}

There is substantial recent literature on combining data-driven learning with reduced-order modeling. One approach focuses on learning reduced dynamics directly from snapshot data; for example, operator inference uses structured regression in reduced coordinates to identify low-dimensional dynamical systems consistent with the underlying reduced model form \cite{PeherstorferWillcox2016,KramerPeherstorferWillcox2024}. An alternative approach is based on learning latent dynamics using autoencoders and nonlinear trial manifolds \cite{FrescaDedeManzoni2021,GeelenBalzanoWrightWillcox2024}. Related hybrid ROM strategies have also combined projection-based models with neural corrections, physics-informed reduced dynamics, or POD--deep-learning prediction architectures \cite{BarnettFarhatMaday2023,ChenWangHesthavenZhang2021,AbadiaHerediaLopezMartinCarroArribasPerezLeClainche2022}.

Our approach is closest in spirit to hybrid methods that preserve a projection-based online solve while using machine learning to improve the reduced representation. However, our method differs from those in much of this literature in two significant ways. First, the learned object is a \emph{time-dependent linear trial subspace}, instead of a latent coordinate update map or a nonlinear decoder manifold. Second, this subspace is predicted directly from the initial condition at the field level by a neural operator, which naturally accommodates functional inputs and outputs. Moreover, using the projection-based online solver acting on the trial subspace, which is data-adaptive and input-conditioned, ensures that this design preserves more physical structure than direct black-box trajectory regression while achieving better compressibility than a fixed linear subspace on dynamical problems.

\section{Problem Formulation}

\subsection{Nonlinear PDE initial value problems}

We consider nonlinear PDE initial value problems of the form
\begin{equation}
    \partial_t u(t,\mathbf{x}) = \mathcal{F}(u(t,\mathbf{x})),
    \quad \mathbf{x}\in\Omega,\quad t\in[0,T],
    \label{eq:pde}
\end{equation}
with initial condition
\begin{equation}
    u(0,\mathbf{x}) = u_0(\mathbf{x}).
    \label{eq:ic}
\end{equation}
Here, \(u(t,\mathbf{x})\) denotes the state variable, \(\Omega \subset \R^d\) is the spatial domain, and \(\mathcal{F}\) is a nonlinear differential operator. 

After spatial discretization, \cref{eq:pde} is approximated by a system of ODEs
\begin{equation}
    \frac{d}{dt}\bm{u}(t) = \bm{F}(\bm{u}(t)),
    \quad \bm{u}(0)=\bm{u}_0,
    \label{eq:semi_discrete}
\end{equation}
where \(\bm{u}(t)\in\mathbb{R}^{N_{\mathrm{dof}}}\), and \(N_{\mathrm{dof}}\) denotes the number of discrete degrees of freedom.

\subsection{Reduced approximation in a moving trial space}

We seek an approximation of the full-order state \(\bm{u}(t)\) in a low-dimensional time-dependent trial space:
\begin{equation}
    \bm{u}(t) \approx \bm{\Phi}(t)\bm{a}(t) + \bm{u}_0,
    \label{eq:rom_ansatz}
\end{equation}
where
\[
\bm{\Phi}(t)= \left[ \phi_1(t),\phi_2(t),\ldots,\phi_r(t) \right]\in Mat_{N_{\mathrm{dof}}\times r}(\R), \quad \bm{a}(t)\in\R^r, \qquad r\ll N_{\mathrm{dof}}.
\]
The columns of the matrix  \(\bm{\Phi}(t)\)  are a time-dependent basis of an $r$-dimensional subspace  $\Lambda(t)\subset\R^{N_{\mathrm{dof}}}$, and \(\bm{a}(t)\) are the reduced coordinates that represent the evolution away from the initial condition. The key question is how to construct \(\bm{\Phi}(t)\) for a previously unseen initial condition \(\bm{u}_0\) in a way that is both inexpensive and sufficiently accurate for reduced simulation.

In standard ROM, one often selects a \emph{fixed} matrix \(\bm{\Phi}\) from global training snapshots and projects \cref{eq:semi_discrete} onto the column space of \(\bm{\Phi}\). Instead, we use a neural operator to predict a matrix \(\bm{\Phi}(t)\) that depends on both the initial condition 
\(\bm{u}_0\) and time $t$.

\section{Method}

\subsection{Overview}

The proposed framework consists of the following stages:
\begin{enumerate}[leftmargin=1.5em]
    \item Generate full-order solution trajectories for a training set of initial conditions.
    \item Construct a time-dependent POD basis for a reduced subspace $\Lambda(t)$ from these trajectories.
    \item Train a neural operator to map an initial condition \(\bm{u}_0\) to the matrix \(\bm{\Phi}(t)\) whose columns are the time-dependent POD basis of $\Lambda(t)$.
    \item For a new initial condition, evaluate the trained neural operator to predict \(\bm{\Phi}(t)\).
    \item Solve a reduced dynamical system for the coordinates \(\bm{a}(t)\) in the predicted moving subspace.
    \item Reconstruct the approximate full state via \(\bm{u}(t)\approx \bm{\Phi}(t)\bm{a}(t) + \bm{u}_0\).
\end{enumerate}

Conceptually, the framework separates \emph{representation learning} from \emph{dynamical evolution}. The neural network is responsible for providing a compact, trajectory-adapted trial space, while the reduced solver enforces the governing dynamics within that trial space.

\subsection{Construction of time-dependent subspace bases}
\label{sec:construction_time_dependent_bases}

For each training trajectory, we assume access to full-order snapshots $u(t_j)$ with $t_j = j \delta t$ and $ j=0,1,\ldots,M$; in our simulations, typically $M \approx 10^3$.

For a given trajectory, we do POD analyses on matrices whose columns are the shifted snapshots \(u_d(t_j)=u(t_j)-u(0)\), at particular times chosen as follows. We take a sliding window time step $\tau = m\delta t $ for some integer $m \geq 1$, and POD time window length $\tau' = m'\delta t$ for some integer $m' \geq r$. In our numerical experiments, we use $m=20$, $m' = 19$ and $r=3,4$. 

Then, we perform POD on the $N \times (m'\! + \! 1)$ snapshot matrix with columns
\[
u_d(t_{jm}),u_d(t_{jm+1}),\ldots,u_d(t_{jm+m'})
\]
where the index j satisfies $j\geq 0$ and $jm+m'\leq M$, and let $\bm\Phi(t_{jm})$ be the $N \times r$ matrix whose columns are the first $r$ vectors in the full POD spatial basis matrix. This yields a sequence of sliding window reduced POD matrices $\bm{\Phi}(t_0),\bm{\Phi}(t_m),\bm{\Phi}(t_{2m}),\ldots$.

Next, we piecewise linearly interpolate this entire reduced POD matrix trajectory with a few temporal breakpoints $t^{\star}_0 (= t_0)< t^{\star}_1 <\ldots < t^{\star}_K$. In the reported experiments, we use five temporal control points for Burgers'
equation and six temporal control points for Fisher--KPP equation. We denote the values of this piecewise linear interpolation matrices at the breakpoints by
\[
\bm{\widetilde{\Phi}}(t^{\star}_{0}), \ldots, \bm{\widetilde{\Phi}}(t^{\star}_{K}).
\]
The data of the matrices $\widetilde{\Phi}$ compared to the original $u(t_j)$ is represented with a compression factor of approximately $\sim \frac{M}{rK}$. In our cases, the compression rate is around 40 to 60. These control point matrices at the temporal breakpoints serve as supervised learning targets for the neural operator introduced in the next section.

\subsection{Neural operator for subspace prediction}

We define a neural operator with training parameters $\theta$
\begin{equation}
    \cG_{\theta}:\bm{u}_0 \mapsto \left(
    \bm{\widetilde{\Phi}}(t^{\star}_{0}), \ldots, \bm{\widetilde{\Phi}}(t^{\star}_{K})
    \right).
    \label{eq:neural_operator_map}
\end{equation}
Let $\bm{\widetilde{\Phi}}(t)$ denote the piecewise linear function on $[t^\star_0,t^\star_K]$ which takes values $\bm{\widetilde{\Phi}}(t^\star_j)$ at $t=t^\star_j$. Then $\bm{\widetilde{\Phi}}(t)$ extends uniquely to a piecewise linear function defined for all $t$, if we impose the additional condition that $\bm{\widetilde{\Phi}}(t)$ is differentiable at $t=t_0^\star$ and $t=t^\star_K$.

In practice, we use a Fourier neural operator architecture because the input and output are naturally represented as grid-based fields, and because FNOs are well-suited to learning mappings between function-valued objects on regular meshes \cite{LiKovachkiAzizzadenesheli2021,KovachkiLiLiuAzizzadenesheliBhattacharyaStuartAnandkumar2023}. The network takes the initial condition field as input and outputs the collection of subspace basis fields at the prescribed temporal breakpoints.

The network is trained to minimize a weighted mixed relative Sobolev--\(L^2\) loss between the predicted and target basis matrices. The discretized Sobolev norm is defined up to first partial derivatives with the \(L^2\) norm. Let
\(\widehat{\bm{\Phi}}_{\theta}^{(n)}\) and \(\bm{\Phi}^{\star (n)}\) denote the predicted and target subspace basis outputs for training sample \(n\), stacked over all temporal breakpoints and basis channels. The training objective is
\begin{equation}
\mathcal{L}(\theta)
=
\frac{1}{N_{\text{train}}}
\sum_{n=1}^{N_{\text{train}}}
\left[
\alpha\,
\frac{
\left\|
W\!\left(\widehat{\bm{\Phi}}_{\theta}^{(n)}-\bm{\Phi}^{\star (n)}\right)
\right\|_{H^{1}}
}{
\left\|
W\bm{\Phi}^{\star (n)}
\right\|_{H^{1}}+\varepsilon
}
+
(1-\alpha)\,
\frac{
\left\|
W\!\left(\widehat{\bm{\Phi}}_{\theta}^{(n)}-\bm{\Phi}^{\star (n)}\right)
\right\|_{L^{2}}
}{
\left\|
W\bm{\Phi}^{\star (n)}
\right\|_{L^{2}}+\varepsilon
}
\right].
\label{eq:basis_loss}
\end{equation}
Here \(W\) denotes a diagonal channel-weighting operator applied along the subspace basis channel dimension, \(\alpha \in [0,1]\) balances the relative \(H^1\) and relative \(L^2\) contributions, and \(\varepsilon>0\) is a small constant for numerical stability. In the experiments, modest channel weights are used to place slightly greater emphasis on the higher-index reduced modes. The use of derivative-enhanced or Sobolev-type training objectives is related to Sobolev training, in which derivative information is incorporated into supervised learning losses \cite{CzarneckiOsinderoJaderberg2017}.

\subsection{Reduced dynamics in a moving subspace}

We start with the ansatz
\begin{equation}
    \bm{u}(t)\approx \bm{\Phi}(t)\bm{a}(t)+\bm{u_0},
\end{equation}
\begin{equation}
    \bm{u}(0)= \bm{u_0},
\end{equation}
so $\bm{a}(0)=\bm{0}$.
Differentiating with respect to time gives
\begin{equation}
    \dot{\bm{u}}(t)\approx \dot{\bm{\Phi}}(t)\bm{a}(t)+\bm{\Phi}(t)\dot{\bm{a}}(t).
    \label{eq:udot_moving_basis}
\end{equation}
Substituting into the semi-discrete full-order system \cref{eq:semi_discrete} gives
\begin{equation}
    \dot{\bm{\Phi}}(t)\bm{a}(t)+\bm{\Phi}(t)\dot{\bm{a}}(t)
    =
    \bm{F}(\bm{\Phi}(t)\bm{a}(t)+\bm{u_0}).
    \label{eq:substitute}
\end{equation}
Projecting onto the trial space with \(\bm{\Phi}(t)^\top\) yields
\begin{equation}
    \bm{\Phi}(t)^\top \bm{\Phi}(t)\dot{\bm{a}}(t)
    +
    \bm{\Phi}(t)^\top\dot{\bm{\Phi}}(t)\bm{a}(t)
    =
    \bm{\Phi}(t)^\top \bm{F}(\bm{\Phi}(t)\bm{a}(t)+\bm{u_0}).
    \label{eq:projected_nonorth}
\end{equation}
Let $\bm{M}(t)=\bm{\Phi}(t)^\top\bm{\Phi}(t);$ then the reduced system becomes
\begin{equation}
    \bm{M}(t)\dot{\bm{a}}(t)
    =
    \bm{\Phi}(t)^\top \bm{F}(\bm{\Phi}(t)\bm{a}(t)+\bm{u_0})
    -
    \bm{\Phi}(t)^\top\dot{\bm{\Phi}}(t)\bm{a}(t).
    \label{eq:moving_rom}
\end{equation}
The second term on the right-hand side is the correction associated with the time variation of the subspace basis. It vanishes for a static subspace basis and is essential in the moving-basis setting. Intuitively, it accounts for the fact that part of the apparent change in the state arises from the motion of the trial space itself rather than from changes in the reduced coordinates. This term is standard in changing-coordinate formulations and must be retained for consistency. Similar correction terms arise naturally in time-dependent subspace basis and dynamically orthogonal formulations \cite{SapsisLermusiaux2009,PatilBabaee2023}.

\subsection{Physics consistency of the projected evolution}

An important conceptual feature of the method is that the neural network does not directly replace the governing dynamics. Instead, it predicts a low-dimensional time-dependent trial space in which the governing equations are approximately enforced by projection. This is more physics-constrained than direct trajectory regression because the online state still results from integrating a reduced dynamical system derived from the PDE. At the same time, the method is more adaptive than standard ROM because the reduced space is conditioned on the particular initial condition field of interest.
The remaining online cost consists of three parts:
\begin{enumerate}[label=(\arabic*)]
    \item inferencing the neural operator to obtain the subspace basis prediction;
    \item building the reduced order model based on the predicted subspace basis;
    \item integrating the reduced system \cref{eq:moving_rom}.
\end{enumerate}

\section{Experimental Setup}

\subsection{Test problems}

\subsubsection{Viscous Burgers' equation}

As a first nonlinear benchmark, we consider the two-dimensional viscous Burgers' equation \cite{Burgers1948} for a vector velocity field \(\mathbf{u}(t,\mathbf{x}) = (u_1(t,\mathbf{x}),u_2(t,\mathbf{x}))^\top\) on the periodic domain \(\Omega=[0,1]^2\):
\begin{equation}
    \partial_t \mathbf{u} + (\mathbf{u}\cdot\nabla)\mathbf{u} = \nu \Delta \mathbf{u},
    \qquad \mathbf{x}=(x,y)\in \Omega,
    \label{eq:burgers}
\end{equation}
with periodic boundary conditions and initial condition
\begin{equation}
    \mathbf{u}(0,\mathbf{x})=\mathbf{u}_0(\mathbf{x}).
\end{equation}
Here \(\nu>0\) denotes the viscosity. We use this problem as a representative nonlinear transport--diffusion system whose solution exhibits evolving spatial structures and nonlinear self-advection, making it a natural test case for time-dependent subspace bases.
In the experiments, the initial velocity fields are generated as smooth random fields on the periodic domain using a band-limited Mat\'ern Gaussian random field construction in Fourier space. This process is explained with more detail in the Appendix. Burgers' equation provides a useful benchmark because it contains both nonlinear advection and diffusive smoothing, and its solutions often exhibit moving coherent structures that are difficult to compress accurately with a single static linear subspace.

\subsubsection{Fisher--KPP Equation}

As a second nonlinear benchmark, we consider the two-dimensional Fisher--KPP reaction--diffusion equation \cite{Fisher1937} on the periodic domain \(\Omega=[0,1]^2\):

\begin{equation}
    \partial_t u = D \Delta u + \rho\, u(1-u),
    \qquad \mathbf{x}\in\Omega,
    \label{eq:fisher_kpp}
\end{equation}
with periodic boundary conditions and initial condition

\begin{equation}
    u(0,\mathbf{x}) = u_0(\mathbf{x}).
\end{equation}
Here $D>0$ is the diffusion constant and $\rho>0$ is the reaction strength.
In the data-generation process, the full-order simulations are run with \(D=0.01\), \(\rho=0.8\), time step \(\Delta t = 10^{-3}\), and final time \(T=1\). The full-order solver uses an IMEX Euler scheme, with diffusion treated implicitly and the logistic reaction term treated explicitly \cite{AscherRuuthSpiteri1997,HundsdorferVerwer2003}.

The initial condition fields are generated from smooth low-frequency random fields on the periodic domain. More specifically, a symmetry-enforced low-frequency Fourier random field is used, with domain length \(L=1\), target standard deviation \(\sigma=0.2\), and Fourier cutoff \(k_{\max}=8\). Each sampled field is then rescaled pointwise to lie in a prescribed interval \(u_0(\mathbf{x}) \in [0.02,0.2]\), which ensures nonnegative and spatially smooth initial data while maintaining diversity across trajectories.

\subsection{Data generation}

For both benchmark problems, the data sets are generated by the same general pipeline. 
First, an initial condition is sampled from the corresponding random initial-condition generator described above. 
The sampled initial condition is then evolved with a high-fidelity full-order solver to produce a reference trajectory
\[
    \left\{ \bm{u}^{(s)}(t_j) \right\}_{j=0}^{M},
    \qquad t_j=j\Delta t,
\]
where \(s\) indexes the trajectory. These full-order trajectories provide the reference snapshots from which the time-dependent reduced POD matrix labels are constructed using the procedure described in \cref{sec:construction_time_dependent_bases}. This leads to supervised training samples of the form
\[
    \left(
    \bm{u}_0^{(s)},\,
    \left\{
    \widetilde{\bm{\Phi}}^{(s)}(t_k^\star)
    \right\}_{k=0}^{K}
    \right),
\]
where \(\bm{u}_0^{(s)}\) is the input initial condition and 
\(\{\widetilde{\bm{\Phi}}^{(s)}(t_k^\star)\}_{k=0}^{K}\) are the reduced-POD matrix control points used as neural-operator targets.

Training data are generated on periodic square grids at multiple spatial resolutions, including \(64^2\), \(128^2\), and \(256^2\). For the Burgers' equation experiments, the training set consists primarily of \(64^2\) trajectories together with additional \(128^2\) trajectories. For the Fisher--KPP equation experiments, the training set consists primarily of \(64^2\) trajectories, supplemented by \(128^2\) trajectories and a smaller number of \(256^2\) trajectories. This multiresolution training strategy encourages the neural operator to learn a resolution-consistent map from initial conditions to reduced subspaces.

For testing, we generate independent full-order trajectories using initial conditions that are not used during training. For each test resolution, we sample \(20\) new random initial conditions from the same initial-condition distribution and evolve them with the corresponding full-order solver. The predicted moving subspaces are then used in the ROM, and the reconstructed ROM trajectories are compared against these independently generated FOM reference trajectories.

In both benchmark problems, spatial coordinate channels are appended to the initial-condition field before it is passed to the neural operator. Therefore, for the Burgers' equation case, the network input consists of the two velocity components together with the \((x,y)\) coordinate channels, while for the Fisher--KPP case, the input consists of the scalar initial field together with the same coordinate channels. The output is the collection of subspace matrices at the prescribed temporal breakpoints.

\subsection{Neural operator training details}

The neural operator is trained to predict time-dependent reduced subspace matrices at a finite number of temporal breakpoints.
For the Burgers' equation case, we use a multi-resolution Fourier Neural Operator with \(5\) layers, \(150\) hidden channels, and Fourier truncation \(n_{\mathrm{modes}}=(32,32)\). The input has \(4\) channels and the output has \(40\) channels. The model contains approximately \(1.23\times 10^8\) trainable parameters.
For the Fisher--KPP equation case, we likewise use a \(5\)-layer Fourier Neural Operator with \(150\) hidden channels, but with a smaller spectral truncation \(n_{\mathrm{modes}}=(16,16)\). The input has \(3\) channels and the output has \(18\) channels. This model contains approximately \(3.27\times 10^7\) trainable parameters.

In both cases, the neural operators are trained with AdamW using learning rate \(10^{-3}\), weight decay \(10^{-4}\), and a cosine annealing learning-rate schedule over \(200\) epochs \cite{LoshchilovHutter2019,LoshchilovHutter2017}. The training is carried out in a multi-resolution fashion by sampling batches from \(64^2\), \(128^2\), and \(256^2\) datasets. The loss used in training is a weighted mixed relative \(H^1\) and relative \(L^2\) loss on the predicted reduced POD matrices channels.

\subsection{Evaluation protocol}

The principal reference is the corresponding full-order model (FOM). For each test case, the trained neural operator first predicts the time-dependent reduced subspace, after which the ROM is integrated forward in time in the predicted moving trial subspace. We then compare the reconstructed ROM trajectory against the full-order trajectory.

\section{Numerical Results}

We evaluate our framework on two nonlinear PDE initial value problems: the two-dimensional viscous Burgers' equation and the two-dimensional Fisher--KPP equation. The purpose of these experiments is to test whether a neural operator can predict a low-dimensional moving reduced subspace that is accurate enough for reduced-order time integration while providing meaningful online speedup over the corresponding full-order solver.

Unless otherwise stated, the reported numerical errors are averaged over the test set. 
For a single test trajectory \(s\), we define the time-dependent mean absolute relative error by
\[
e^{(s)}(t)
=
\frac{1}{N_{\mathrm{dof}}}
\sum_{\ell=1}^{N_{\mathrm{dof}}}
\frac{
\left|
\widehat{u}^{(s)}_{\ell}(t)
-
u_{\mathrm{FOM},\ell}^{(s)}(t)
\right|
}{
\left|
u_{\mathrm{FOM},\ell}^{(s)}(t)
\right|+\varepsilon
},
\]
where \(N_{\mathrm{dof}}=N^2\) for the Fisher--KPP equation, \(N_{\mathrm{dof}}=2N^2\) for Burgers' equation, \(u_{\mathrm{FOM}}^{(s)}\) is the full-order reference solution, 
\(\widehat{u}^{(s)}\) is our method's prediction, \(N\) is the number of grid points in each spatial direction, and \(\varepsilon=10^{-12}\) is a small constant used to avoid division by zero.

For a test set containing \(N_{\mathrm{test}}\) independently sampled trajectories, we report the average error that is computed as
\[
\overline{e}(t)
=
\frac{1}{N_{\mathrm{test}}}
\sum_{s=1}^{N_{\mathrm{test}}}
e^{(s)}(t).
\]
Thus, the plotted quantity is the test-set average of the single-trajectory error \(e^{(s)}(t)\) at each time. For vector-valued Burgers' equation solutions, the error is computed over the velocity field components. Runtime is reported as the wall-clock speedup
\[
S = \frac{T_{\mathrm{FOM}}}{T_{\mathrm{Ours}}},
\]
where values larger than one indicate that our method is faster than the full-order solver.

\subsection{Burgers' equation results}

\begin{figure}[t]
    \centering

    \includegraphics[width=0.75\textwidth]{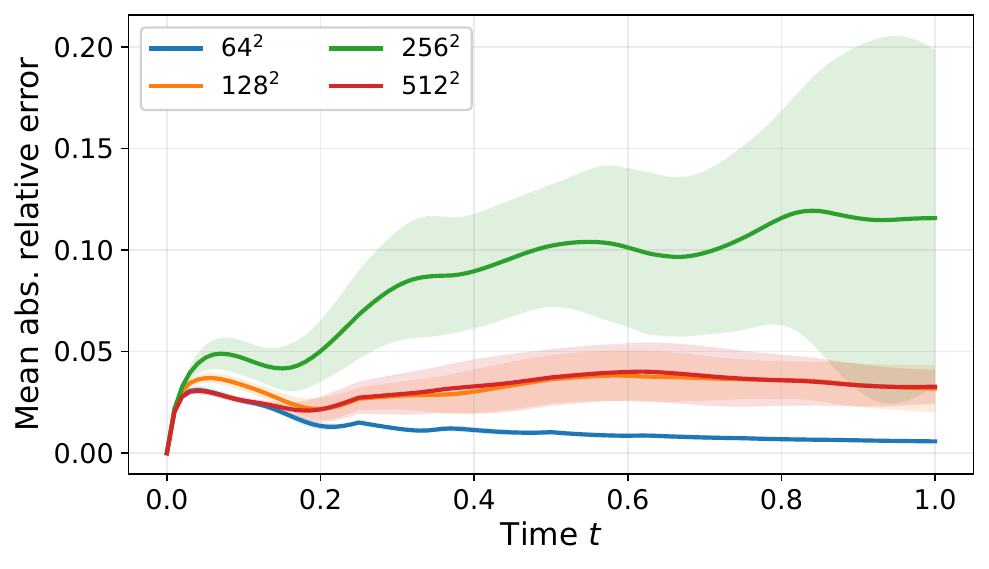}

    \vspace{0.5em}

    \includegraphics[width=0.48\textwidth]{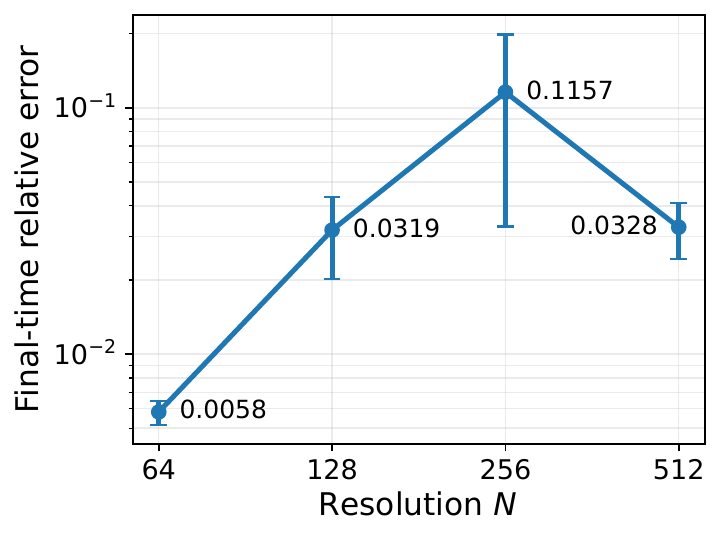}
    \includegraphics[width=0.48\textwidth]{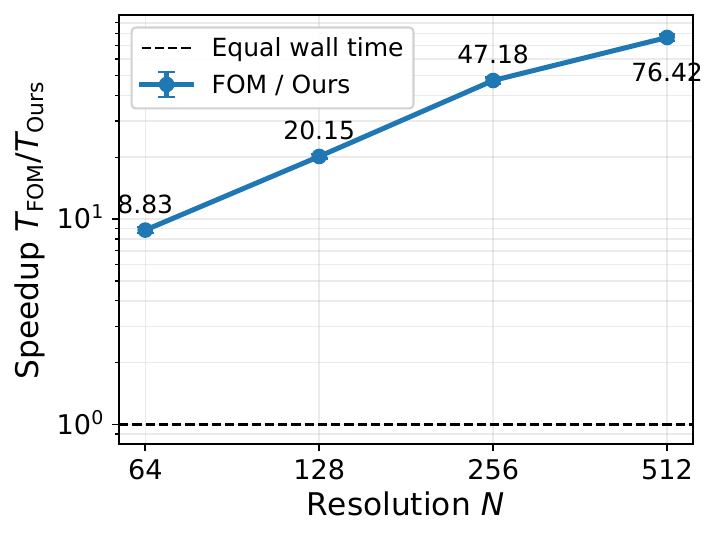}

    \caption{
    Burgers' equation results.
    Top: test-set mean absolute relative error over time.
    Bottom left: final-time mean absolute relative error versus spatial resolution.
    Bottom right: wall-clock speedup \(T_{\mathrm{FOM}}/T_{\mathrm{Ours}}\).
    }
    \label{fig:burgers_results}
\end{figure}

Figure~\ref{fig:burgers_results} reports the Burgers' equation results across spatial resolutions \(64^2\), \(128^2\), \(256^2\), and \(512^2\). The error curves show that our method remains accurate over the full time interval for most resolutions. The final-time mean absolute relative errors are approximately \(5.8\times 10^{-3}\), \(3.19\times 10^{-2}\), \(1.157\times 10^{-1}\), and \(3.28\times 10^{-2}\) for \(64^2\), \(128^2\), \(256^2\), and \(512^2\), respectively. Thus, our method achieves errors below \(4\times 10^{-2}\) at \(64^2\), \(128^2\), and \(512^2\), while the \(256^2\) case exhibits noticeably larger error and variance.

The non-monotone behavior at \(256^2\) suggests a resolution-dependent error source in the current learned subspace prediction or reduced model operator assembly. Nevertheless, our method remains stable over the simulation horizon, and the error stays at the order of \(10^{-1}\) even in the worst tested case.

The computational gain increases strongly with resolution. The measured wall-clock speedups are approximately \(8.83\times\), \(20.15\times\), \(47.18\times\), and \(76.42\times\) for \(64^2\), \(128^2\), \(256^2\), and \(512^2\), respectively. This trend is consistent with the purpose of the method: the full-order solver cost grows rapidly with the spatial grid size, while our method's cost remains low.

\FloatBarrier

\subsection{Fisher--KPP equation results}

\begin{figure}[t]
    \centering

    \includegraphics[width=0.75\textwidth]{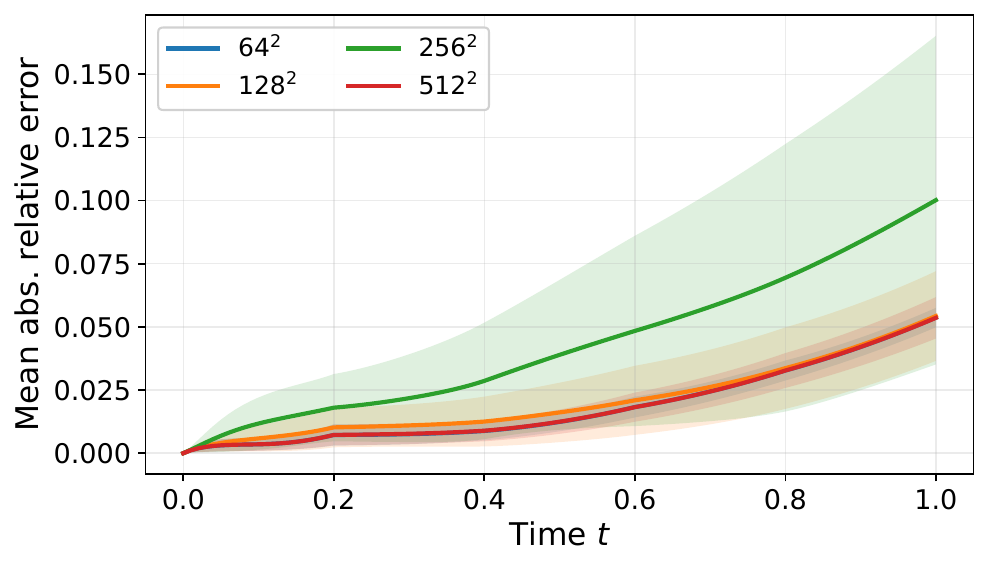}

    \vspace{0.5em}

    \includegraphics[width=0.48\textwidth]{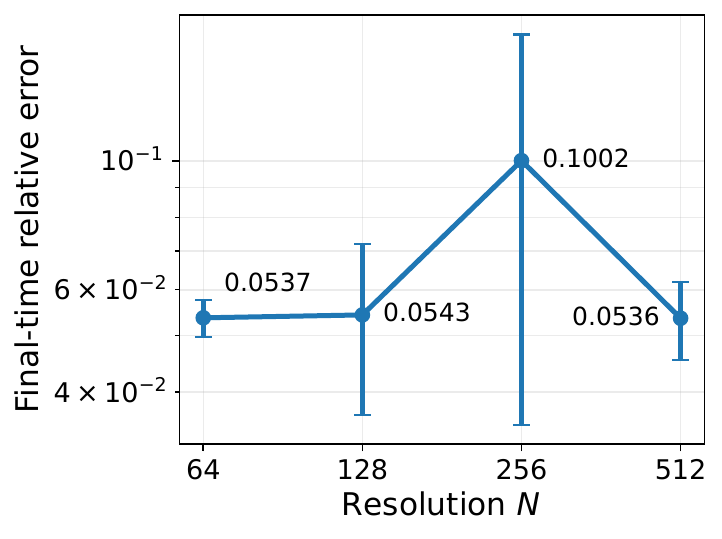}
    \includegraphics[width=0.48\textwidth]{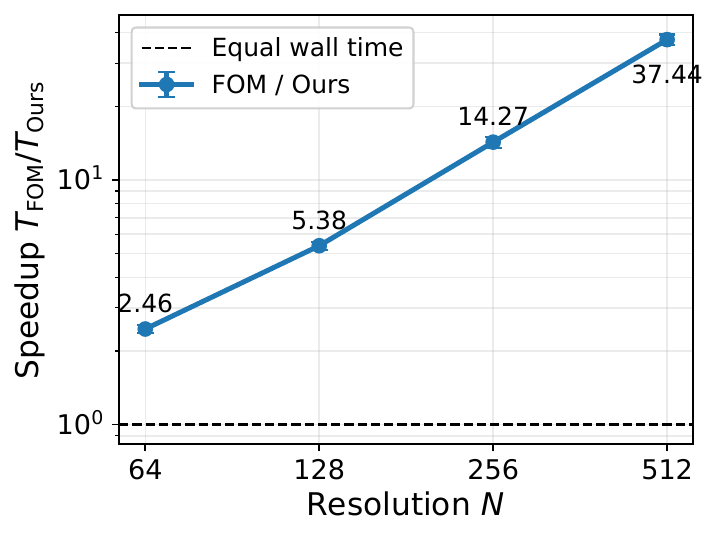}

    \caption{
    Fisher--KPP equation results.
    Top: test-set mean absolute relative error over time.
    Bottom left: final-time mean absolute relative error as a function of spatial resolution.
    Bottom right: wall-clock speedup \(T_{\mathrm{FOM}}/T_{\mathrm{Ours}}\).
    }
    \label{fig:kpp_results}
\end{figure}

Figure~\ref{fig:kpp_results} reports the corresponding results for the Fisher--KPP equation. In contrast to Burgers' equation, the Fisher--KPP equation has a nonlinear reaction term instead of a nonlinear advection term.

The final-time mean absolute relative errors are approximately \(5.37\times 10^{-2}\), \(5.43\times 10^{-2}\), \(1.002\times 10^{-1}\), and \(5.36\times 10^{-2}\) for \(64^2\), \(128^2\), \(256^2\), and \(512^2\), respectively. The \(64^2\), \(128^2\), and \(512^2\) cases have nearly identical final-time errors around \(5.4\times 10^{-2}\). The \(256^2\) case has the largest error and the largest variance, with final-time error around \(10^{-1}\).

The time-dependent error curves show gradual error growth over the time horizon, which is expected for a long-time evolution problem. The relative error remains around \(10^{-1}\) across all tested resolutions. The results indicate that the predicted moving basis captures the dominant reaction--diffusion dynamics sufficiently well for reduced simulation.

The speedup is again significant and increases with resolution. The measured wall-clock speedups are approximately \(2.46\times\), \(5.38\times\), \(14.27\times\), and \(37.44\times\) for \(64^2\), \(128^2\), \(256^2\), and \(512^2\), respectively. The speedup is smaller than in the Burgers' equation case at comparable resolutions. Still, the same scaling trend is observed: our method becomes more advantageous as the full-order spatial grid becomes larger.

\FloatBarrier

\subsection{Summary of numerical results}

Taken together, the Burgers' equation and Fisher--KPP equation experiments support the central idea of this paper: instead of learning the full order solution trajectory directly, it is often sufficient to learn a low-dimensional time-dependent trial subspace and then evolve the reduced coordinates by solving a projected dynamical system in that subspace. This hybrid strategy retains the physics-based structure of a ROM solve while using a neural operator to adapt the reduced subspace to the specific initial condition. In both examples, this leads to accurate trajectories together with meaningful online speedup relative to the corresponding full-order simulation. 

\section{Discussion}

\subsection{When the method is likely to work well}

The proposed framework is likely to be most effective when the solution trajectories have only a few dominant spatial modes, but which are strongly initial condition-dependent. This includes settings where the dominant structures evolve or translate over time, making a fixed global basis inefficient, yet where the governing dynamics still admit a compact representation in a moving linear space. Burgers' equation and Fisher--KPP equation both fit this picture to a meaningful extent.

The method is especially appealing in many-query regimes where the same PDE must be solved repeatedly for varying initial data. In such settings, the offline training cost can be amortized, and the online speedup becomes operationally relevant.

\subsection{Advantages}

This framework offers several advantages:
\begin{itemize}[leftmargin=1.5em]
    \item It uses a PDE-based online ROM solver to model the time evolution rather than a black-box predictor.
    \item It allows the reduced trial space to depend on the initial condition, improving adaptivity relative to a fixed basis ROM.
    \item It leverages neural operators for field-to-field prediction, enabling generalization across different spatial resolutions.
    \item It is modular: improvements in basis extraction, neural operator architectures, temporal interpolation, or hyper-reduction can all be incorporated without changing the overall method.
\end{itemize}

\subsection{Limitations}

Several limitations remain.
First, although the reduced subspace is low-dimensional, evaluating nonpolynomial or nonlinear terms in \cref{eq:moving_rom} may still require full-order operations unless hyper-reduction is incorporated, for example through EIM-, DEIM-, GNAT-, or empirical-cubature-style approximations \cite{BarraultMadayNguyenPatera2004,ChaturantabutSorensen2010,CarlbergFarhatCortialAmsallem2013}. Thus the present method improves efficiency but does not yet realize the full online complexity reductions available in advanced nonlinear ROM pipelines.

Second, basis prediction is subject to ambiguities in sign, ordering, and rotation, especially when multiple modes have similar importance. A simple matrix-norm training loss may not optimally respect the underlying subspace geometry. This suggests using losses or optimization formulations defined on Stiefel or Grassmann manifolds \cite{EdelmanAriasSmith1998,AbsilMahonySepulchre2008}.

Third, the present numerical study, while encouraging, is still limited in scope. Broader validation on more challenging PDE families and stronger baselines would be needed to fully establish the strengths and limitations of this approach.

\subsection{Future work}

Several directions appear especially promising:
\begin{itemize}[leftmargin=1.5em]
    \item integrating hyper-reduction, such as EIM-, DEIM-, GNAT-, or empirical-cubature-style ideas, into the moving-basis ROM;
    \item designing subspace-specific training losses that are invariant to basis sign and within-subspace rotations;
    \item learning temporally coherent basis parameterizations rather than predicting independent basis snapshots at breakpoints;
    \item incorporating structure-preserving constraints, especially for conservative or Hamiltonian systems;
    \item extending the framework to parameterized PDE families beyond initial-condition variation.
\end{itemize}

\section{Conclusion}

We have presented a hybrid framework for nonlinear PDE initial value problems in which a neural operator predicts a low-dimensional time-dependent reduced subspace from the initial condition, and the reduced coordinates are then evolved by solving a projected reduced dynamical system in this moving trial space. The method combines two complementary ingredients: expressive function-to-function learning for basis prediction, and a physics-based reduced solver for dynamical evolution.

Numerical experiments on the viscous Burgers' equation and the Fisher--KPP equation show that the proposed method achieves substantial wall-clock speedups that increase with spatial resolution, reaching up to \(76.42\times\) for Burgers' equation and \(37.44\times\) for Fisher--KPP equation, while maintaining final-time mean absolute relative errors on the order of \(10^{-2}\) to \(10^{-1}\) across tested cases. These results support the view that learning an adaptive reduced trial space, rather than the full trajectory directly, is a viable strategy for data-assisted simulation of nonlinear PDEs.

More broadly, the method highlights a promising interface between neural operators and reduced-order modeling. Instead of treating machine learning and numerical model reduction as competing approaches, it uses learning to improve the reduced representation while preserving a projection-based dynamical core. We expect that richer basis representations, stronger subspace-aware learning objectives, and additional hyper-reduction will further improve the method's efficiency and robustness.

\appendix

\section{Full-order discretization}
\label{app:fom_discretization}

Both benchmark problems are solved on the periodic domain
\[
    \Omega=[0,1]^2
\]
using a uniform square grid
\[
    x_i=i\Delta x,\qquad y_j=j\Delta y,
    \qquad i,j=0,\ldots,N-1,\qquad \Delta x=\Delta y=\frac{1}{N}.
\]
All finite-difference stencils use periodic wrapping.

For diffusion, we use the fourth-order centered second derivative. For example,
\[
    D_{xx} f_{i,j}
    =
    \frac{
    -f_{i-2,j}
    +16f_{i-1,j}
    -30f_{i,j}
    +16f_{i+1,j}
    -f_{i+2,j}
    }{12\Delta x^2},
\]
with an analogous formula for \(D_{yy}\). The discrete Laplacian is
\[
    \Delta_h f = D_{xx}f+D_{yy}f.
\]

\subsection{Viscous Burgers' equation}

For the two-dimensional viscous Burgers' equation,
\[
    \partial_t \mathbf{u}
    +
    (\mathbf{u}\cdot\nabla)\mathbf{u}
    =
    \nu\Delta\mathbf{u},
    \qquad
    \mathbf{u}=(u,v)^\top,
\]
we use
\[
    \nu=0.05,\qquad \Delta t=10^{-3},\qquad T=1.
\]
The full-order solver uses an IMEX Euler scheme \cite{AscherRuuthSpiteri1997,HundsdorferVerwer2003}. The nonlinear advection term is treated explicitly with second-order upwind derivatives, while diffusion is treated implicitly \cite{LeVeque2002}. One time step is
\begin{align}
    (I-\Delta t\,\nu\Delta_h)u^{n+1}
    &=
    u^n
    -
    \Delta t
    \left(
    u^nD_x^{\rm up}u^n
    +
    v^nD_y^{\rm up}u^n
    \right),
    \\
    (I-\Delta t\,\nu\Delta_h)v^{n+1}
    &=
    v^n
    -
    \Delta t
    \left(
    u^nD_x^{\rm up}v^n
    +
    v^nD_y^{\rm up}v^n
    \right).
\end{align}
The implicit Helmholtz systems are solved using a matrix-free preconditioned conjugate-gradient iteration \cite{HestenesStiefel1952,Saad2003}.

\subsection{Fisher--KPP equation}

For the Fisher--KPP equation,
\[
    \partial_t u
    =
    D\Delta u+\rho u(1-u),
\]
we use
\[
    D=0.01,\qquad \rho=0.8,\qquad \Delta t=10^{-3},\qquad T=1.
\]
The full-order solver also uses an IMEX Euler scheme, with implicit diffusion and explicit reaction \cite{AscherRuuthSpiteri1997,HundsdorferVerwer2003}:
\[
    (I-\Delta t\,D\Delta_h)u^{n+1}
    =
    u^n+\Delta t\,\rho u^n(1-u^n).
\]
The resulting Helmholtz system is solved by the same matrix-free preconditioned conjugate-gradient iteration \cite{HestenesStiefel1952,Saad2003}.

\subsection{Matrix-free solution of implicit diffusion systems}
In both solvers, the implicit diffusion step requires systems of the form
\[
    (I-\alpha\Delta_h)x=b,
\]
with \(\alpha=\nu\Delta t\) for Burgers' equation and \(\alpha=D\Delta t\) for Fisher--KPP equation. These systems are solved matrix-free: the operator is applied as
\[
    H_\alpha p = p-\alpha\Delta_h p
\]
without explicitly assembling a sparse matrix. A Jacobi preconditioner is used with
\[
    P_\alpha^{-1}r
    =
    \frac{r}{
    1+2.5\alpha(\Delta x^{-2}+\Delta y^{-2})
    },
\]
which is the inverse diagonal of \(I-\alpha\Delta_h\) for the fourth-order centered Laplacian. The previous time-step solution is used as the initial guess, and the iteration is stopped when the relative residual is below \(10^{-10}\), with a maximum of \(200\) iterations.

\section{Initial-condition generation}
\label{app:ic_generation}

\subsection{Viscous Burgers' equation}

The construction follows the standard Mat\'ern/Gaussian-random-field viewpoint \cite{Matern1960,RasmussenWilliams2006,LindgrenRueLindstrom2011}. For Burgers' equation, the two velocity components are sampled independently
from the same periodic band-limited Mat\'ern-type Gaussian random field
distribution. Let
\[
    \mathbf{k}=(k_x,k_y)\in\mathbb{Z}^2,
    \qquad
    |\mathbf{k}|=\sqrt{k_x^2+k_y^2}.
\]
The real-valued fluctuation field is
\[
    \eta(x,y)
    = C
    \sum_{|k_x|,|k_y|\le k_{max}}
    A_{\mathbf{k}}\xi_{\mathbf{k}}
    e^{2\pi i(k_xx+k_yy)},
\]
where the coefficient \(\xi_{\mathbf{k}}\)'s real and imaginary parts are chosen according to a Gaussian distribution with mean 0 and variance 1, subject to the constraint \(\xi_{\mathbf{k}} =\bar  \xi_{\mathbf{-k}}\), scaled by the factor
\[
    A_{\mathbf{k}}
    =
    \left[
    1+\left(2\pi\ell|\mathbf{k}|\right)^2
    \right]^{-(\nu_m+1)/2},
\]
and the normalization factor $C$ is chosen so that $\eta$ has specified standard deviation
\[
    C
    \Bigl\{
    \sum_{1\le{|k_x|,|k_y|\le k_{max}}}
    A_{\mathbf{k}}^2
    |\xi_{\mathbf{k}}|^2
    \Bigr\}^{1/2}
    =\sigma
    .
\]
The Burgers' equation initial condition is then
\[
    u_0(x,y)=1+\eta_1(x,y),
    \qquad
    v_0(x,y)=1+\eta_2(x,y),
\]
where \(\eta_1\) and \(\eta_2\) are independent samples.

The parameters used in the reported Burgers' equation simulations are
\[
    L=1,\qquad
    \sigma=0.2,\qquad
    \ell=0.05,\qquad
    \nu_m=0.5,\qquad
    k_{\max}=8.
\]

\subsection{Fisher--KPP equation}

For the Fisher--KPP equation, the scalar initial condition is generated from a smooth low-frequency cosine random field. Specifically,
\[
    s(x,y)
    =
    C \sum_{m,n}
    c_{mn}
    \cos\!\left(\frac{2\pi m x}{L}\right)
    \cos\!\left(\frac{2\pi n y}{L}\right),
\]
with modes \(1\leq m,n\leq k_{\max}\), subject to the constraint $c_{mn}=c_{nm}$ where $c_{mn}$ is drawn from a Gaussian distribution with mean $0$ and standard deviation $(m^2+n^2)^{-3/2}$. $C$ is chosen so that
\[
    \frac{C}{2L}
    \Bigl\{
    \sum_{m,n}
    c_{mn}^2
    \Bigr\}^{1/2}
    =\sigma
    .
\]

We rescale $s(x,y)$
according to
\[
    u_0(x,y)
    =
    u_{\min}
    +
    (u_{\max}-u_{\min})
    \frac{
    s(x,y)-\min_{\Omega}s
    }{
    \max_{\Omega}s-\min_{\Omega}s
    },
\]
so that the resulting field $u_0(x,y)$ has range $[u_{min}, u_{max}]$; in our simulation we set $[u_{min}, u_{max}] = [0.02, 0.2]$. The remaining parameters used in the reported Fisher--KPP equation simulations are
\[
    L=1,\qquad
    \sigma=0.2,\qquad
    k_{\max}=8.
\]

\section{Time-dependent POD construction}
\label{app:td_pod_labels}

For each full-order trajectory
\[
    \{\bm{u}(t_j)\}_{j=0}^{M},
    \qquad t_j=j\Delta t,
\]
local reduced POD matrices are computed from sliding windows of shifted trajectory snapshots
\[
    \bm{u}_d(t_j)=\bm{u}(t_j)-\bm{u}_0.
\]
This is consistent with the subspace approximation
\[
    \bm{u}(t)\approx \bm{u}_0+\bm{\Phi}(t)\bm{a}(t).
\]

A window of \(m_w=20\) saved snapshots and stride \(m_s=20\) is used. For a window beginning at index \(q\), define
\[
    X_q
    =
    \begin{bmatrix}
    \bm{u}_d(t_q)^\top\\
    \bm{u}_d(t_{q+1})^\top\\
    \vdots\\
    \bm{u}_d(t_{q+m_w-1})^\top
    \end{bmatrix}
    \in \mathbb{R}^{m_w\times N_{\rm dof}},
\]
where \(N_{\rm dof}=2N^2\) for Burgers' equation and \(N_{\rm dof}=N^2\) for the Fisher--KPP equation. The SVD
\[
    X_q=U_q\Sigma_qV_q^\top
\]
is computed, and the leading right singular vectors define the local subspace. The retained dimensions are \(r=4\) per velocity component for Burgers' equation and \(r=3\) for the Fisher--KPP equation.

The resulting sequence of local reduced POD matrices is compressed in time by fitting a degree-one B-spline, equivalently a piecewise linear function, to a small set of temporal control points. The Burgers' case uses five temporal control points, while the Fisher--KPP case uses six temporal control points. These control-point value matrices are used as the supervised targets for neural-operator training.

\section{Neural-operator training}
\label{app:fno_training}

The neural operator maps the initial condition to the time-dependent basis control point value matrices. Coordinate channels are appended to the inputs. Thus, the Burgers' input channels are
\(
    (u_0,v_0,x,y),
\)
and the Fisher--KPP input channels are
\(
    (u_0,x,y).
\)
The coordinate channels use cell-centered coordinates.

For Burgers' equation, the output contains \(5\times2\times4=40\)
channels: five temporal control points, two physical components, and four retained modes per component; for the Fisher--KPP equation, the  model uses \(6\times1\times3=18\)
output channels. The complete neural-operator architecture and training settings are
summarized in \cref{tab:app_fno_training}.

\begin{table}[h]
\centering
\caption{Neural-operator architecture and training settings.}
\label{tab:app_fno_training}
\begin{tabular}{lll}
\toprule
Quantity & Burgers' equation & Fisher--KPP equation \\
\midrule
Input channels & \(4\) & \(3\) \\
Output channels & \(40\) & \(18\) \\
FNO layers & \(5\) & \(5\) \\
Hidden channels & \(150\) & \(150\) \\
Fourier modes & \((32,32)\) & \((16,16)\) \\
Retained reduced modes & \(r=4\) per component & \(r=3\) \\
Optimizer & AdamW & AdamW \\
Learning rate & \(10^{-3}\) & \(10^{-3}\) \\
Weight decay & \(10^{-4}\) & \(10^{-4}\) \\
Epochs & \(200\) & \(200\) \\
Scheduler & Cosine annealing & Cosine annealing \\
Training split & \(90\%/10\%\) & \(90\%/10\%\) \\
\bottomrule
\end{tabular}
\end{table}

The weighted mixed relative \(H^1\)--\(L^2\) loss used in the main text has equal weighting between the two terms.
The reduced-mode channel weights are
\(
    (1.0,1.2,1.2,1.6)
\)
for Burgers' equation and
\(
    (1.0,1.2,1.2)
\)
for Fisher--KPP equation.

\section{Moving-subspace ROM implementation}
\label{app:rom_implementation}

The ROM approximation is
\[
    \bm{u}(t)\approx \bm{u}_0+\bm{\Phi}(t)\bm{a}(t),
    \qquad
    \bm{a}(0)=0.
\]
Substitution into the semi-discrete full-order system and Galerkin projection gives
\[
    \bm{M}(t)\dot{\bm{a}}(t)
    =
    \bm{\Phi}(t)^\top
    \bm{F}\!\left(\bm{u}_0+\bm{\Phi}(t)\bm{a}(t)\right)
    -
    \bm{\Phi}(t)^\top \dot{\bm{\Phi}}(t)\bm{a}(t),
\]
where
\[
    \bm{M}(t)=\bm{\Phi}(t)^\top\bm{\Phi}(t).
\]
The predicted subspace is represented piecewise linearly in time, so
\(\dot{\bm{\Phi}}(t)\) is piecewise constant. The reduced ODE is integrated
using a BDF solver with absolute and relative tolerances \(10^{-6}\)
\cite{HairerWanner1996}.

All spatial inner products below are approximated by uniform random sampling of \(N_s\)
grid indices \(j\):
\[
    \langle f,g\rangle
    \approx
    \frac{N_xN_y}{N_s}
    \sum_{j} f_j g_j.
\]
We use \(N_s=4000\) for both Burgers' equation and the Fisher--KPP equation.
The current implementation does not include hyper-reduction, such as EIM, DEIM,
or GNAT \cite{BarraultMadayNguyenPatera2004,ChaturantabutSorensen2010,CarlbergFarhatCortialAmsallem2013}.

\subsection{Reduced operators for Burgers' equation}

For Burgers' equation, write the initial condition as
\[
    \bm{u}_0=(u_0,v_0)^\top.
\]
The moving-subspace approximation is
\[
    u(t,x,y)
    \approx
    u_0(x,y)+\sum_{i=1}^{r} a_i(t)\phi_i(t,x,y),
\]
\[
    v(t,x,y)
    \approx
    v_0(x,y)+\sum_{i=1}^{r} b_i(t)\psi_i(t,x,y).
\]
The reduced state is
\[
    \bm{y}(t)
    =
    (a_1,\ldots,a_r,b_1,\ldots,b_r)^\top.
\]
The full-order Burgers' equations are
\[
    \partial_t u
    =
    \nu\Delta u
    -
    u\,\partial_x u
    -
    v\,\partial_y u,
\]
\[
    \partial_t v
    =
    \nu\Delta v
    -
    u\,\partial_x v
    -
    v\,\partial_y v.
\]
Define the mass matrices
\[
    A^u_{ij}(t)
    =
    \langle \phi_i,\phi_j\rangle,
    \qquad
    A^v_{ij}(t)
    =
    \langle \psi_i,\psi_j\rangle.
\]
The constant projected terms are
\[
    c^u_i(t)
    =
    \left\langle
    \phi_i,\,
    \nu\Delta u_0
    -
    u_0\,\partial_x u_0
    -
    v_0\,\partial_y u_0
    \right\rangle,
\]
\[
    c^v_i(t)
    =
    \left\langle
    \psi_i,\,
    \nu\Delta v_0
    -
    u_0\,\partial_x v_0
    -
    v_0\,\partial_y v_0
    \right\rangle.
\]
The linear operators in the reduced \(u\)-equation are
\[
    L^{uu}_{ij}(t)
    =
    \left\langle
    \phi_i,\,
    \nu\Delta\phi_j
    -
    u_0\,\partial_x\phi_j
    -
    v_0\,\partial_y\phi_j
    -
    \phi_j\,\partial_x u_0
    -
    \partial_t\phi_j
    \right\rangle,
\]
\[
    L^{uv}_{ij}(t)
    =
    -
    \left\langle
    \phi_i,\,
    \psi_j\,\partial_y u_0
    \right\rangle.
\]
The linear operators in the reduced \(v\)-equation are
\[
    L^{vu}_{ij}(t)
    =
    -
    \left\langle
    \psi_i,\,
    \phi_j\,\partial_x v_0
    \right\rangle,
\]
\[
    L^{vv}_{ij}(t)
    =
    \left\langle
    \psi_i,\,
    \nu\Delta\psi_j
    -
    u_0\,\partial_x\psi_j
    -
    v_0\,\partial_y\psi_j
    -
    \psi_j\,\partial_y v_0
    -
    \partial_t\psi_j
    \right\rangle.
\]
The terms \(\partial_t\phi_j\) and \(\partial_t\psi_j\) are the
moving-subspace correction terms.

The quadratic terms in the reduced \(u\)-equation have coefficients
\[
    Q^{uu}_{kij}(t)
    =
    -
    \left\langle
    \phi_k,\,
    \phi_i\,\partial_x\phi_j
    \right\rangle,
\]
\[
    Q^{uv}_{kij}(t)
    =
    -
    \left\langle
    \phi_k,\,
    \psi_j\,\partial_y\phi_i
    \right\rangle.
\]
The quadratic terms in the reduced \(v\)-equation have coefficients
\[
    Q^{vu}_{\ell ij}(t)
    =
    -
    \left\langle
    \psi_\ell,\,
    \phi_j\,\partial_x\psi_i
    \right\rangle,
\]
\[
    Q^{vv}_{\ell ij}(t)
    =
    -
    \left\langle
    \psi_\ell,\,
    \psi_j\,\partial_y\psi_i
    \right\rangle.
\]

With these definitions, the reduced Burgers system is
\[
    \sum_{i=1}^{r} A^u_{ki}\dot a_i
    =
    c^u_k
    +
    \sum_{i=1}^{r} L^{uu}_{ki}a_i
    +
    \sum_{i=1}^{r} L^{uv}_{ki}b_i
    +
    \sum_{i,j=1}^{r} Q^{uu}_{kij}a_i a_j
    +
    \sum_{i,j=1}^{r} Q^{uv}_{kij}a_i b_j,
\]
\[
    \sum_{i=1}^{r} A^v_{\ell i}\dot b_i
    =
    c^v_\ell
    +
    \sum_{i=1}^{r} L^{vu}_{\ell i}a_i
    +
    \sum_{i=1}^{r} L^{vv}_{\ell i}b_i
    +
    \sum_{i,j=1}^{r} Q^{vu}_{\ell ij}b_i a_j
    +
    \sum_{i,j=1}^{r} Q^{vv}_{\ell ij}b_i b_j.
\]
Equivalently, in block form,
\[
    \bm{A}(t)\dot{\bm{y}}(t)
    =
    \bm{C}(t)
    +
    \bm{L}(t)\bm{y}(t)
    +
    \bm{Q}(t)(\bm{y}(t))
\]
where
\[
    \bm{A}(t)
    =
    \begin{bmatrix}
    A^u(t) & 0\\
    0 & A^v(t)
    \end{bmatrix},
    \qquad
    \bm{C}(t)
    =
    \begin{bmatrix}
    c^u(t)\\
    c^v(t)
    \end{bmatrix}
\]
and
\[
    \bm{L}(t)
    =
    \begin{bmatrix}
    L^{uu}(t) & L^{uv}(t)\\
    L^{vu}(t) & L^{vv}(t)
    \end{bmatrix}.
\]

\subsection{Reduced operators for the Fisher--KPP equation}

For the Fisher--KPP equation, the moving-basis approximation is
\[
    u(t,x,y)
    \approx
    u_0(x,y)+\sum_{i=1}^{r}a_i(t)\phi_i(t,x,y).
\]
The full-order equation is
\[
    \partial_t u
    =
    D\Delta u+\rho u(1-u)
    =
    D\Delta u+\rho u-\rho u^2.
\]
Substituting the shifted approximation gives
\[
    u
    \approx
    u_0+\sum_i a_i\phi_i.
\]
The constant term is
\[
    D\Delta u_0+\rho u_0-\rho u_0^2,
\]
the linear term in \(a_i\) is
\[
    D\Delta\phi_i+\rho\phi_i-2\rho u_0\phi_i,
\]
the quadratic term in \(a_i\) is
\[
    -\rho \phi_i\phi_j.
\]
Including the moving-subspace correction gives the ROM operators
\[
    M_{ij}(t)
    =
    \langle \phi_i,\phi_j\rangle,
\]
\[
    c_i(t)
    =
    \left\langle
    \phi_i,\,
    D\Delta u_0+\rho u_0-\rho u_0^2
    \right\rangle,
\]
\[
    L_{ij}(t)
    =
    \left\langle
    \phi_i,\,
    D\Delta\phi_j+\rho\phi_j-2\rho u_0\phi_j-\partial_t\phi_j
    \right\rangle,
\]
\[
    Q_{kij}(t)
    =
    \left\langle
    \phi_k,\,
    -\rho\phi_i\phi_j
    \right\rangle.
\]
Thus the Fisher--KPP reduced system is
\[
    \sum_{i=1}^{r} M_{ki}(t)\dot a_i(t)
    =
    c_k(t)
    +
    \sum_{i=1}^{r} L_{ki}(t)a_i(t)
    +
    \sum_{i,j=1}^{r} Q_{kij}(t)a_i(t)a_j(t),
    \qquad k=1,\ldots,r.
\]
In vector form,
\[
    \bm{M}(t)\dot{\bm{a}}(t)
    =
    \bm{c}(t)
    +
    \bm{L}(t)\bm{a}(t)
    +
    \bm{Q}(t)(\bm{a}(t)).
\]

\section{Evaluation protocol}
\label{app:evaluation_protocol}

For each tested resolution
\(N = {64,128,256,512}\);
we generate \(20\) independent test trajectories. The reported time-dependent error is the mean absolute relative error over the spatial degrees of freedom. For one test trajectory \(s\), it is
\[
e^{(s)}(t)
=
\frac{1}{N_{\rm dof}}
\sum_{i=1}^{N_{\rm dof}}
\frac{
\left|
\widehat{u}^{(s)}_i(t)
-
u^{(s)}_{{\rm FOM},i}(t)
\right|
}{
\left|
u^{(s)}_{{\rm FOM},i}(t)
\right|+\varepsilon
},
\qquad
\varepsilon=10^{-12}.
\]
For the Fisher--KPP equation, \(N_{\rm dof}=N^2\). For Burgers' equation, both velocity components are flattened into one vector, so \(N_{\rm dof}=2N^2\). The plotted error is the test-set average
\[
    \overline{e}(t)
    =
    \frac{1}{N_{\rm test}}
    \sum_{s=1}^{N_{\rm test}}e^{(s)}(t).
\]

The reported speedup is
\[
    S=\frac{T_{\rm FOM}}{T_{\rm Ours}},
\]
where \(T_{\rm Ours}\) includes neural-operator inference, reduced subspace interpolation, reduced-operator assembly, and reduced dynamics time integration. The final-time errors and measured speedup factors are summarized in
\cref{tab:app_summary}.

\begin{table}[h]
\centering
\caption{Final-time mean absolute relative error and wall-clock speedup.}
\label{tab:app_summary}
\begin{tabular}{lccc}
\toprule
Problem & Resolution & Final-time error & Speedup factor \\
\midrule
Burgers & \(64^2\)  & \(5.8\times10^{-3}\)   & \(8.83\) \\
Burgers & \(128^2\) & \(3.19\times10^{-2}\)  & \(20.15\) \\
Burgers & \(256^2\) & \(1.157\times10^{-1}\) & \(47.18\) \\
Burgers & \(512^2\) & \(3.28\times10^{-2}\)  & \(76.42\) \\
Fisher--KPP & \(64^2\)  & \(5.37\times10^{-2}\)  & \(2.46\) \\
Fisher--KPP & \(128^2\) & \(5.43\times10^{-2}\)  & \(5.38\) \\
Fisher--KPP & \(256^2\) & \(1.002\times10^{-1}\) & \(14.27\) \\
Fisher--KPP & \(512^2\) & \(5.36\times10^{-2}\)  & \(37.44\) \\
\bottomrule
\end{tabular}
\end{table}

\section*{Declaration of competing interest}
The authors declare that they do not have any financial or personal relationships with other individuals or organizations that inappropriately influenced or biased this work. 

\section*{Acknowledgements}
This research was partially supported by NSF grant DMS--1910303.

\section*{Data and code availability}
Data and code will be shared in a GitHub repository upon acceptance of the paper.

\section*{Declaration of generative AI use}
During the preparation of this work, the authors used ChatGPT to transcribe parts of an early draft. Subsequently, the tool's output was substantially revised by the authors who take full responsibility for the content of the published article.

\bibliographystyle{unsrtnat}
\bibliography{references}

@article{Sirovich1987,
  author  = {Sirovich, Lawrence},
  title   = {Turbulence and the Dynamics of Coherent Structures. {Part I}: Coherent Structures},
  journal = {Quarterly of Applied Mathematics},
  volume  = {45},
  number  = {3},
  pages   = {561--571},
  year    = {1987},
  doi     = {10.1090/qam/910462}
}

@article{BennerGugercinWillcox2015,
  author  = {Benner, Peter and Gugercin, Serkan and Willcox, Karen},
  title   = {A Survey of Projection-Based Model Reduction Methods for Parametric Dynamical Systems},
  journal = {SIAM Review},
  volume  = {57},
  number  = {4},
  pages   = {483--531},
  year    = {2015},
  doi     = {10.1137/130932715}
}

@article{HaasdonkOhlberger2008,
  author  = {Haasdonk, Bernard and Ohlberger, Mario},
  title   = {Reduced Basis Method for Finite Volume Approximations of Parametrized Linear Evolution Equations},
  journal = {ESAIM: Mathematical Modelling and Numerical Analysis},
  volume  = {42},
  number  = {2},
  pages   = {277--302},
  year    = {2008},
  doi     = {10.1051/m2an:2008001}
}

@article{HesthavenPagliantiniRozza2022,
  author  = {Hesthaven, Jan S. and Pagliantini, Cecilia and Rozza, Gianluigi},
  title   = {Reduced Basis Methods for Time-Dependent Problems},
  journal = {Acta Numerica},
  volume  = {31},
  pages   = {265--345},
  year    = {2022},
  doi     = {10.1017/S0962492922000058}
}

@article{ReissSchulzeSesterhennMehrmann2018,
  author  = {Reiss, Julius and Schulze, Philipp and Sesterhenn, J{\"o}rn and Mehrmann, Volker},
  title   = {The Shifted Proper Orthogonal Decomposition: A Mode Decomposition for Multiple Transport Phenomena},
  journal = {SIAM Journal on Scientific Computing},
  volume  = {40},
  number  = {3},
  pages   = {A1322--A1344},
  year    = {2018},
  doi     = {10.1137/17M1140571}
}

@article{BarnettFarhatMaday2023,
  author  = {Barnett, Joshua L. and Farhat, Charbel and Maday, Yvon},
  title   = {Neural-Network-Augmented Projection-Based Model Order Reduction for Mitigating the {Kolmogorov} Barrier to Reducibility of {CFD} Models},
  journal = {Journal of Computational Physics},
  volume  = {492},
  pages   = {112420},
  year    = {2023},
  doi     = {10.1016/j.jcp.2023.112420}
}

@article{LuJinPangKarniadakis2021,
  author  = {Lu, Lu and Jin, Pengzhan and Pang, Guofei and Zhang, Zhongqiang and Karniadakis, George Em},
  title   = {Learning Nonlinear Operators via {DeepONet} Based on the Universal Approximation Theorem of Operators},
  journal = {Nature Machine Intelligence},
  volume  = {3},
  number  = {3},
  pages   = {218--229},
  year    = {2021},
  doi     = {10.1038/s42256-021-00302-5}
}

@inproceedings{LiKovachkiAzizzadenesheli2021,
  author    = {Li, Zongyi and Kovachki, Nikola and Azizzadenesheli, Kamyar and Liu, Burigede and Bhattacharya, Kaushik and Stuart, Andrew and Anandkumar, Anima},
  title     = {Fourier Neural Operator for Parametric Partial Differential Equations},
  booktitle = {International Conference on Learning Representations},
  year      = {2021},
  url       = {https://openreview.net/forum?id=c8P9NQVtmnO}
}

@article{KovachkiLiLiuAzizzadenesheliBhattacharyaStuartAnandkumar2023,
  author  = {Kovachki, Nikola and Li, Zongyi and Liu, Burigede and Azizzadenesheli, Kamyar and Bhattacharya, Kaushik and Stuart, Andrew M. and Anandkumar, Anima},
  title   = {Neural Operator: Learning Maps Between Function Spaces with Applications to {PDEs}},
  journal = {Journal of Machine Learning Research},
  volume  = {24},
  number  = {89},
  pages   = {1--97},
  year    = {2023},
  url     = {https://jmlr.org/papers/v24/21-1524.html}
}

@article{PeherstorferWillcox2015,
  author  = {Peherstorfer, Benjamin and Willcox, Karen},
  title   = {Online Adaptive Model Reduction for Nonlinear Systems via Low-Rank Updates},
  journal = {SIAM Journal on Scientific Computing},
  volume  = {37},
  number  = {4},
  pages   = {A2123--A2150},
  year    = {2015},
  doi     = {10.1137/140989169}
}

@article{SapsisLermusiaux2009,
  author  = {Sapsis, Themistoklis P. and Lermusiaux, Pierre F. J.},
  title   = {Dynamically Orthogonal Field Equations for Continuous Stochastic Dynamical Systems},
  journal = {Physica D: Nonlinear Phenomena},
  volume  = {238},
  number  = {23--24},
  pages   = {2347--2360},
  year    = {2009},
  doi     = {10.1016/j.physd.2009.09.017}
}

@article{PatilBabaee2023,
  author  = {Patil, Prerna and Babaee, Hessam},
  title   = {Reduced-Order Modeling with Time-Dependent Bases for {PDEs} with Stochastic Boundary Conditions},
  journal = {SIAM/ASA Journal on Uncertainty Quantification},
  volume  = {11},
  number  = {3},
  pages   = {727--756},
  year    = {2023},
  doi     = {10.1137/21M1468097}
}

@article{DonelloPalkarNaderiDelReyFernandezBabaee2023,
  author  = {Donello, Michael and Palkar, Gaurav and Naderi, Mohammad Hossein and Del Rey Fern{\'a}ndez, David C. and Babaee, Hessam},
  title   = {Oblique Projection for Scalable Rank-Adaptive Reduced-Order Modelling of Nonlinear Stochastic Partial Differential Equations with Time-Dependent Bases},
  journal = {Proceedings of the Royal Society A: Mathematical, Physical and Engineering Sciences},
  volume  = {479},
  number  = {2278},
  pages   = {20230320},
  year    = {2023},
  doi     = {10.1098/rspa.2023.0320}
}

@article{PeherstorferWillcox2016,
  author  = {Peherstorfer, Benjamin and Willcox, Karen},
  title   = {Data-Driven Operator Inference for Nonintrusive Projection-Based Model Reduction},
  journal = {Computer Methods in Applied Mechanics and Engineering},
  volume  = {306},
  pages   = {196--215},
  year    = {2016},
  doi     = {10.1016/j.cma.2016.03.025}
}

@article{KramerPeherstorferWillcox2024,
  author  = {Kramer, Boris and Peherstorfer, Benjamin and Willcox, Karen E.},
  title   = {Learning Nonlinear Reduced Models from Data with Operator Inference},
  journal = {Annual Review of Fluid Mechanics},
  volume  = {56},
  number  = {1},
  pages   = {521--548},
  year    = {2024},
  doi     = {10.1146/annurev-fluid-121021-025220}
}

@article{FrescaDedeManzoni2021,
  author  = {Fresca, Stefania and Ded{\`e}, Luca and Manzoni, Andrea},
  title   = {A Comprehensive Deep Learning-Based Approach to Reduced Order Modeling of Nonlinear Time-Dependent Parametrized {PDEs}},
  journal = {Journal of Scientific Computing},
  volume  = {87},
  number  = {2},
  pages   = {61},
  year    = {2021},
  doi     = {10.1007/s10915-021-01462-7}
}

@article{GeelenBalzanoWrightWillcox2024,
  author  = {Geelen, Rudy and Balzano, Laura and Wright, Stephen and Willcox, Karen},
  title   = {Learning Physics-Based Reduced-Order Models from Data Using Nonlinear Manifolds},
  journal = {Chaos: An Interdisciplinary Journal of Nonlinear Science},
  volume  = {34},
  number  = {3},
  pages   = {033122},
  year    = {2024},
  doi     = {10.1063/5.0170105}
}

@article{ChenWangHesthavenZhang2021,
  author  = {Chen, Wenqian and Wang, Qian and Hesthaven, Jan S. and Zhang, Chuhua},
  title   = {Physics-Informed Machine Learning for Reduced-Order Modeling of Nonlinear Problems},
  journal = {Journal of Computational Physics},
  volume  = {446},
  pages   = {110666},
  year    = {2021},
  doi     = {10.1016/j.jcp.2021.110666}
}

@article{AbadiaHerediaLopezMartinCarroArribasPerezLeClainche2022,
  author  = {Abad{\'i}a-Heredia, Rodrigo and L{\'o}pez-Mart{\'i}n, Manuel and Carro, Bel{\'e}n and Arribas, Juan Ignacio and P{\'e}rez, Jos{\'e} Manuel and Le Clainche, Soledad},
  title   = {A Predictive Hybrid Reduced Order Model Based on Proper Orthogonal Decomposition Combined with Deep Learning Architectures},
  journal = {Expert Systems with Applications},
  volume  = {187},
  pages   = {115910},
  year    = {2022},
  doi     = {10.1016/j.eswa.2021.115910}
}

@inproceedings{CzarneckiOsinderoJaderberg2017,
  author    = {Czarnecki, Wojciech M. and Osindero, Simon and Jaderberg, Max and Swirszcz, Grzegorz and Pascanu, Razvan},
  title     = {Sobolev Training for Neural Networks},
  booktitle = {Advances in Neural Information Processing Systems},
  volume    = {30},
  year      = {2017},
  url       = {https://papers.nips.cc/paper_files/paper/2017/hash/758a06618c69880a6cee5314ee42d52f-Abstract.html}
}

@incollection{Burgers1948,
  author    = {Burgers, Johannes M.},
  title     = {A Mathematical Model Illustrating the Theory of Turbulence},
  booktitle = {Advances in Applied Mechanics},
  volume    = {1},
  pages     = {171--199},
  publisher = {Academic Press},
  year      = {1948},
  doi       = {10.1016/S0065-2156(08)70100-5}
}

@book{LeVeque2002,
  author    = {LeVeque, Randall J.},
  title     = {Finite Volume Methods for Hyperbolic Problems},
  publisher = {Cambridge University Press},
  address   = {Cambridge},
  year      = {2002},
  doi       = {10.1017/CBO9780511791253}
}

@article{Fisher1937,
  author  = {Fisher, Ronald A.},
  title   = {The Wave of Advance of Advantageous Genes},
  journal = {Annals of Eugenics},
  volume  = {7},
  number  = {4},
  pages   = {355--369},
  year    = {1937},
  doi     = {10.1111/j.1469-1809.1937.tb02153.x}
}

@article{AscherRuuthSpiteri1997,
  author  = {Ascher, Uri M. and Ruuth, Steven J. and Spiteri, Raymond J.},
  title   = {Implicit-Explicit {Runge--Kutta} Methods for Time-Dependent Partial Differential Equations},
  journal = {Applied Numerical Mathematics},
  volume  = {25},
  number  = {2--3},
  pages   = {151--167},
  year    = {1997},
  doi     = {10.1016/S0168-9274(97)00056-1}
}

@book{HundsdorferVerwer2003,
  author    = {Hundsdorfer, Willem and Verwer, Jan G.},
  title     = {Numerical Solution of Time-Dependent Advection-Diffusion-Reaction Equations},
  publisher = {Springer},
  address   = {Berlin},
  year      = {2003},
  doi       = {10.1007/978-3-662-09017-6}
}

@article{HestenesStiefel1952,
  author  = {Hestenes, Magnus R. and Stiefel, Eduard},
  title   = {Methods of Conjugate Gradients for Solving Linear Systems},
  journal = {Journal of Research of the National Bureau of Standards},
  volume  = {49},
  number  = {6},
  pages   = {409--436},
  year    = {1952},
  doi     = {10.6028/jres.049.044}
}

@book{Saad2003,
  author    = {Saad, Yousef},
  title     = {Iterative Methods for Sparse Linear Systems},
  publisher = {Society for Industrial and Applied Mathematics},
  address   = {Philadelphia, PA},
  edition   = {2nd},
  year      = {2003},
  doi       = {10.1137/1.9780898718003}
}

@article{Matern1960,
  author  = {Mat{\'e}rn, Bertil},
  title   = {Spatial Variation: Stochastic Models and Their Application to Some Problems in Forest Surveys and Other Sampling Investigations},
  journal = {Meddelanden fr{\aa}n Statens Skogsforskningsinstitut},
  volume  = {49},
  number  = {5},
  pages   = {1--144},
  year    = {1960}
}

@book{RasmussenWilliams2006,
  author    = {Rasmussen, Carl Edward and Williams, Christopher K. I.},
  title     = {Gaussian Processes for Machine Learning},
  publisher = {MIT Press},
  address   = {Cambridge, MA},
  year      = {2006},
  url       = {https://gaussianprocess.org/gpml/}
}

@article{LindgrenRueLindstrom2011,
  author  = {Lindgren, Finn and Rue, H{\aa}vard and Lindstr{\"o}m, Johan},
  title   = {An Explicit Link Between {Gaussian} Fields and {Gaussian Markov} Random Fields: The Stochastic Partial Differential Equation Approach},
  journal = {Journal of the Royal Statistical Society Series B: Statistical Methodology},
  volume  = {73},
  number  = {4},
  pages   = {423--498},
  year    = {2011},
  doi     = {10.1111/j.1467-9868.2011.00777.x}
}

@article{BarraultMadayNguyenPatera2004,
  author  = {Barrault, Maxime and Maday, Yvon and Nguyen, Ngoc Cuong and Patera, Anthony T.},
  title   = {An Empirical Interpolation Method: Application to Efficient Reduced-Basis Discretization of Partial Differential Equations},
  journal = {Comptes Rendus Mathematique},
  volume  = {339},
  number  = {9},
  pages   = {667--672},
  year    = {2004},
  doi     = {10.1016/j.crma.2004.08.006}
}

@article{ChaturantabutSorensen2010,
  author  = {Chaturantabut, Saifon and Sorensen, Danny C.},
  title   = {Nonlinear Model Reduction via Discrete Empirical Interpolation},
  journal = {SIAM Journal on Scientific Computing},
  volume  = {32},
  number  = {5},
  pages   = {2737--2764},
  year    = {2010},
  doi     = {10.1137/090766498}
}

@article{CarlbergFarhatCortialAmsallem2013,
  author  = {Carlberg, Kevin and Farhat, Charbel and Cortial, Julien and Amsallem, David},
  title   = {The {GNAT} Method for Nonlinear Model Reduction: Effective Implementation and Application to Computational Fluid Dynamics and Turbulent Flows},
  journal = {Journal of Computational Physics},
  volume  = {242},
  pages   = {623--647},
  year    = {2013},
  doi     = {10.1016/j.jcp.2013.02.028}
}

@article{EdelmanAriasSmith1998,
  author  = {Edelman, Alan and Arias, Tom{\'a}s A. and Smith, Steven T.},
  title   = {The Geometry of Algorithms with Orthogonality Constraints},
  journal = {SIAM Journal on Matrix Analysis and Applications},
  volume  = {20},
  number  = {2},
  pages   = {303--353},
  year    = {1998},
  doi     = {10.1137/S0895479895290954}
}

@book{AbsilMahonySepulchre2008,
  author    = {Absil, P.-A. and Mahony, Robert and Sepulchre, Rodolphe},
  title     = {Optimization Algorithms on Matrix Manifolds},
  publisher = {Princeton University Press},
  address   = {Princeton, NJ},
  year      = {2008},
  doi       = {10.1515/9781400830244}
}

@book{HairerWanner1996,
  author    = {Hairer, Ernst and Wanner, Gerhard},
  title     = {Solving Ordinary Differential Equations {II}: Stiff and Differential-Algebraic Problems},
  series    = {Springer Series in Computational Mathematics},
  volume    = {14},
  publisher = {Springer},
  address   = {Berlin, Heidelberg},
  edition   = {2nd},
  year      = {1996},
  doi       = {10.1007/978-3-642-05221-7}
}

@inproceedings{LoshchilovHutter2019,
  author    = {Loshchilov, Ilya and Hutter, Frank},
  title     = {Decoupled Weight Decay Regularization},
  booktitle = {International Conference on Learning Representations},
  year      = {2019},
  url       = {https://openreview.net/forum?id=Bkg6RiCqY7}
}

@inproceedings{LoshchilovHutter2017,
  author    = {Loshchilov, Ilya and Hutter, Frank},
  title     = {{SGDR}: Stochastic Gradient Descent with Warm Restarts},
  booktitle = {International Conference on Learning Representations},
  year      = {2017},
  url       = {https://openreview.net/forum?id=Skq89Scxx}
}

\end{document}